\documentclass[a4paper,10pt]{article}
\usepackage[utf8]{inputenc}
\usepackage{graphicx}
\usepackage{xfrac}
\usepackage{amsmath,amsfonts,amssymb}
\usepackage{subcaption}
\usepackage{color}
\usepackage{bm}
\usepackage[left=4cm,right=4cm]{geometry}
\usepackage{slashed}
\usepackage{cite}
\usepackage{csquotes}
\usepackage{tabularx}
\usepackage{authblk}
\usepackage[normalem]{ulem}
\usepackage[colorlinks=true, urlcolor=blue, linkcolor=blue, citecolor=blue]{hyperref}
\usepackage[toc,page]{appendix}


\newcommand{\KaTie}{Ka\hspace{-0.2ex}Tie}

\title{\bf On the  normalization of unintegrated parton densities}
\author[1]{Benjamin Guiot\thanks{benjamin.guiot@usm.cl}}
\affil[1]{Departamento de F\'isica, Universidad T\'ecnica Federico Santa Mar\'ia; Casilla 110-V, Valparaiso, Chile}

\date{}

\begin{document}

\maketitle

\begin{abstract}
Recently, the definitions of the Kimber-Martin-Ryskin-Watt (KMRW) unintegrated parton densities (UPDFs) have been discussed by several groups. In the first part of this manuscript, we remind the issues encountered with these definitions and discuss the proposed solutions. In our opinion, none of these solutions is fully satisfactory. We observe that these issues seem to be related to the normalization condition where the UPDFs are related to the collinear PDFs by an integration over transverse momentum cut off by the factorization scale. Then, we build a modified version of the angular-ordering KMRW UPDFs, obeying the normalization condition with the transverse momentum integrated up to infinity and show that the usual issues are absent. 
\end{abstract}
\newpage

\tableofcontents

\section{Introduction}
Transverse-momentum-dependent parton distribution functions (TMD PDFs) play a central role in the description of two-scale observables, and provide valuable information on the proton internal dynamics. The TMD PFDs appear in the TMD factorization \cite{JiMaYu1,JiMaYu2,BaBoDi,col}, proven to all orders for several processes, and in the high-energy factorization \cite{semihard1,ktFac3,ktFac,ktFac2,semihard2}. In the latter, TMD PDFs are called unintegrated PDFs (UPDFs), and the cross section is obtained by the convolution of these functions with off-shell cross sections. For hadron-hadron collisions, we have
\begin{multline}
 \frac{d\sigma}{dx_1dx_2d^2p_t}(s,x_1,x_2,p_t)=\sum_{a,b}\int^{k_{t,\text{max}}^2}_0 d^2k_{1t}d^2k_{2t} F_{a/h}(x_1,k_{1t};\mu)\\
 \times F_{b/h}(x_2,k_{2t};\mu)\hat{\sigma}(x_1x_2s,k_{1t},k_{2t},p_t;\mu), \label{ktfac}
\end{multline}
with  $F_{a/h}$ the UPDFs for flavor $a=q,\bar{q},g$, $x$ the fraction of the hadron longitudinal momentum carried by the parton, $k_t$ the initial-parton transverse momentum, and $\mu$ the factorization scale. The upper limit of integration is typically given by $k_{t,\text{max}}^2\sim s$, with $\sqrt{s}$ the center-of-mass energy of the collision.\\

A convenient and widely-used set of UPDFs is obtained with the Kimber-Martin-Ryskin-Watt (KMRW) approach \cite{kmr,wmr}, whose starting point is the relation
\begin{equation}
    \widetilde{f}_a(x,\mu)=\int^{\mu^2}_0 F_a(x,k_t;\mu)dk_t^2, \label{undens}
\end{equation}
with $\widetilde{f}_a(x,\mu)=xf_a(x,\mu)$ and $f_a(x,\mu)$ the collinear PDFs. The final result for the definitions of the UPDFs is given in \cite{wmr}.\footnote{In the following, we mention only Ref.~\cite{wmr} where the first work \cite{kmr} has been improved.} The differential form of these UPDFs reads
\begin{eqnarray}
F_a(x,k_t;\mu)&=&\frac{\partial}{\partial k_t^2}\left[T_a(k_t,\mu)\widetilde{f}_a(x,k_t)\right], \quad k_t \geq \mu_0\label{def1}\\
F_a(x,k_t;\mu^2)&=&\frac{1}{\mu_0^2} T_a(\mu_0,\mu)\widetilde{f}_a(x,\mu_0), \quad k_t<\mu_0,\label{conkt}
\end{eqnarray}
where $\mu_0\sim 1$ GeV and $T_a$ is the Sudakov form factor
\begin{equation}
    T_a(k_t,\mu)=\exp\left(- \int_{k_t^2}^{\mu^2}\frac{dq^2}{q^2}\frac{\alpha_s(q^2)}{2\pi}\sum_b\int_{z_{ab}^{\text{min}}(q,\mu)}^{z_{ab}^{\text{max}}(q,\mu)} dz \, z\hat{P}_{ba}(z) \right).\label{suda}
\end{equation}
$\hat{P}_{ba}$ are the unregularized splitting functions. Some of these functions have divergences which are regularized by the cutoffs $z_{ab}^{\text{min}}$ and $z_{ab}^{\text{max}}$, with $z_{ab}^{\text{min}}$ either equal to $0$ or $1-z_{ab}^{\text{max}}$ \cite{wmr}. Note that in the above definition, the cutoffs should not depend on $k_t$ \cite{BGS,Guiot:2019vsm}. In that case, using
\begin{equation}
    \frac{\partial T_a(k_t,\mu)}{\partial \ln k_t^2}=\frac{\alpha_s(k_t^2)}{2\pi}T_a(k_t,\mu)\sum_{b}\int_{z_{ab}^{\text{min}}(k_t,\mu)}^{z_{ab}^{\text{max}}(k_t,\mu)} dz\,z\hat{P}_{ba}(z),\label{derta}
\end{equation}
and the cutoff-dependent Dokshitzer-Gribov-Lipatov-Altarelli-Parisi (DGLAP) equation
\begin{equation}
\frac{\partial \widetilde{f}_a(x,\mu,z_{ab}^{\text{max}})}{\partial \ln\mu^2}=\sum_{b}\frac{\alpha_s(\mu^2)}{2\pi}\left[\int_x^{z_{ab}^{\text{max}}}\hat{P}_{ab}(z)\widetilde{f}_b\left(\frac{x}{z},\mu\right)dz-\widetilde{f}_a(x,\mu)\int_{z_{ab}^{\text{min}}}^{z_{ab}^{\text{max}}}z\hat{P}_{ba}(z)dz \right], \label{wmrdglap}
\end{equation}
we can write Eq.~(\ref{def1}) in its integral form
\begin{equation}
F_a(x,k_t;\mu)=\frac{\alpha_s(k_t^2)}{2\pi}\frac{T_a(k_t,\mu)}{k_t^2}\sum_{b}\int_x^{z_{ab}^{\text{max}}(k_t,\mu)}dz\hat{P}_{ab}(z)\widetilde{f}_b\left(\frac{x}{z},k_t\right).\label{def2}
\end{equation}
Note that in Eq.~(\ref{def2}), and contrary to Eq.~(\ref{suda}), the cutoffs depend on $k_t$. This expression is equal to those given in \cite{wmr}, albeit a more general notation for the cutoffs. With the angular-ordering (AO) cutoffs, we have
\begin{align}
    z_{gg}^{\text{max}}&=z_{qq}^{\text{max}}=\frac{\mu}{\mu+k_t}, \quad z_{gq}^{\text{max}}=z_{qg}^{\text{max}}=1, \label{ao1}\\
    z_{gg}^{\text{min}}&=1-z_{gg}^{\text{max}}, \quad z_{qq}^{\text{min}}=z_{gq}^{\text{min}}=z_{qg}^{\text{min}}=0. \label{ao2}
\end{align}
In Sec.~\ref{secdef}, we remind the reader of some issues related to the integral definition of the KMRW UPDFs, and present in Sec.~\ref{secsol} two proposed solutions. We argue that none of these solutions are fully satisfying. In Sec.~\ref{secpb}, we observe that UDPFs obeying the normalization condition Eq.~(\ref{norinf}) instead of Eq.~(\ref{undens}) seem free of these issues. Our main result is given in Sec.~\ref{secmkmrw}, where we build KMRW-like UPDFs with normalization (\ref{norinf}). We show that these modified KMRW (mKMRW) UPDFs are free of any of the issues listed Sec.~\ref{seciss}.

\section{Troubles with the KMRW formalism \label{sectrou}}

\subsection{Integral definition \label{secdef}}
A first issue with the integral definition is that the normalization condition (\ref{undens}) is not always fulfilled. A second issue is that the two definitions, Eqs.~(\ref{def1}) and (\ref{def2}), are, in general, not equivalent. These problems have already been discussed in the literature \cite{BGS,Guiot:2019vsm,Nefedov:2020ugj,Valeshabadi:2021smo}.\\

The first reason explaining the nonequivalence of Eqs.~(\ref{def1}) and (\ref{def2}) is that the Sudakov factor does not fulfill the condition on $k_t$, see Eq.~(\ref{suda}) and the discussion below. Indeed, the condition $T_a=1$ if $k_t>\mu$ is imposed to avoid a Sadukov factor larger than 1. Then, the actual expression for the Sudakov factor is
\begin{equation}
    T_a(k_t,\mu)=\Theta(\mu^2-k_t^2)\exp\left(- \int_{k_t^2}^{\mu^2}\frac{dq^2}{q^2}\frac{\alpha_s(q^2)}{2\pi}\sum_b\int_{z_{ab}^{\text{min}}(q,\mu)}^{z_{ab}^{\text{max}}(q,\mu)} dz \, z\hat{P}_{ba}(z) \right)+\Theta(k_t^2-\mu^2),\label{sudamod}
\end{equation}
which has an additional dependence on $k_t$ compared to Eq.~(\ref{suda}). Starting from the definition Eq.~(\ref{def1}) and using the cutoff-dependent DGLAP equation (\ref{wmrdglap}), we arrive at \cite{Guiot:2019vsm}
\begin{multline}
F_a(x,k_t;\mu)=\frac{\alpha_s(k_t^2)}{2\pi k_t^2}\left(T_a(k_t,\mu)\sum_{b}\int_x^{z_{ab}^{\text{max}}(k_t,\mu)}dz\hat{P}_{ab}(z)\widetilde{f}_b\left(\frac{x}{z},k_t\right)\right.\\
\left. -\Theta(k_t^2-\mu^2)\widetilde{f}_a(x,k_t)\sum_{b}\int_{z_{ab}^{\text{min}}(k_t,\mu)}^{z_{ab}^{\text{max}}(k_t,\mu)} dz\,z\hat{P}_{ba}(z)\right). \label{def3}
\end{multline}
Compared to Eq.~(\ref{def2}), we see that the correct expression has a new term accompanied by a step function.\\

The cutoff $z_{ab}^{\text{max}}$ should be close to 1 such that $(1-z_{ab}^{\text{max}})\sim \mathcal{O}(\Lambda_{QCD}/\mu)$, with $\mu$ of the order of the hard scale. Indeed, the collinear PDFs in Eq.~(\ref{def1}) obey the exact DGLAP equation. However, we used the cutoff-dependent DGLAP equation to obtain Eqs.~(\ref{def2}) and (\ref{def3}). It is known that the cutoff-dependent and exact DGLAP equation are equal, up to corrections of order $\mathcal{O}(1-z_{ab}^{\text{max}})$; see, for instance, \cite{Hautmann:2017fcj}. Consequently, a second reason for the nonequivalence of the two definitions is the use of a cutoff outside the region $(1-z_{ab}^{\text{max}})\sim \mathcal{O}(\Lambda_{QCD}/\mu)$, which is the case for the KMRW UPDFs. This last point is directly related to the violation of the normalization condition (\ref{undens}) by the integral definition. 

\subsection{Complete list of issues \label{seciss}}
\begin{figure}[h]
\begin{center}
 \includegraphics[width=25pc]{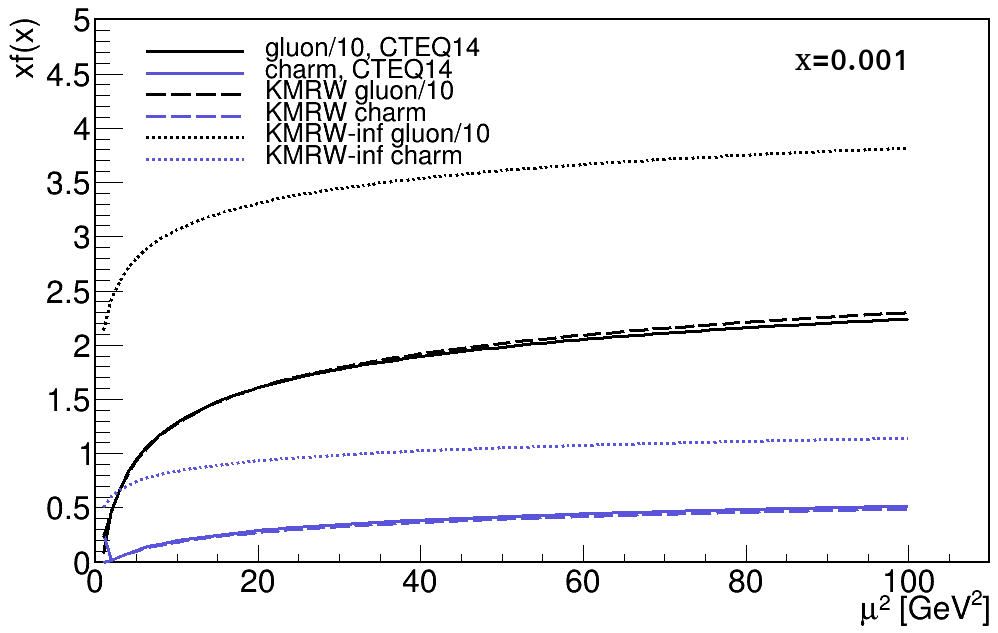}
\end{center}
\caption{\label{wmrpdf} Full and dashed lines correspond to the left- and right-hand sides of Eq.~(\ref{undens}), respectively. The dotted lines correspond to the case where the same UPDFs are integrated up to infinity instead of $\mu^2$.}
\end{figure}

\begin{figure}[h]
\begin{center}
 \includegraphics[width=25pc]{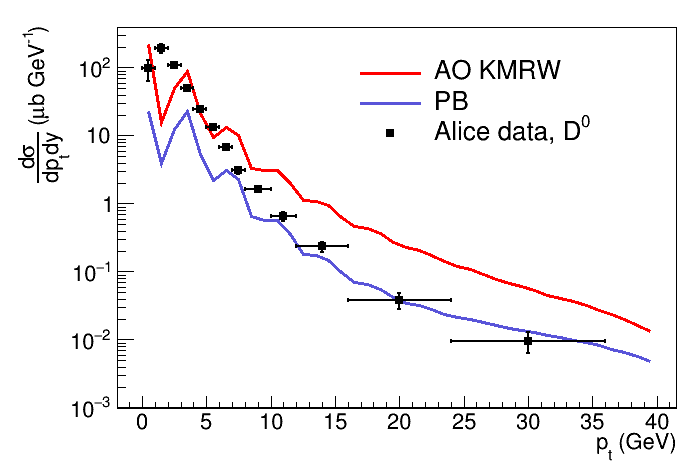}
\end{center}
\caption{\label{cgcg} $d\sigma(cg \to D^0+X)/dp_t$ compared to ALICE data \cite{ALICE:2017olh} at $\sqrt{s}=7$ TeV. The red line has been obtained with the AO KMRW UPDFs presented in Fig. \ref{wmrpdf}, the blue line with the parton-branching UPDFs.}
\end{figure}

Another issue related to the normalization condition (\ref{undens}) is that the tail ($k_t>\mu$) of the distribution is unconstrained but does contribute to the cross section. It opens the possibility for two distributions (built at the same order and with the same scheme) obeying Eq.~(\ref{undens}) to give a quite different result for the cross section. For instance, the AO KMRW UPDFs strongly overestimates the $D$-meson cross section \cite{Guiot:2019vsm,Guiot:2018kfy}. To show this explicitly, we have implemented the AO KMRW UPDFs following exactly Ref. \cite{wmr}. We check our implementation by plotting the left- and right-hand sides of Eq.~(\ref{undens}) in Fig. \ref{wmrpdf}.
We observe a good agreement between the full and dashed lines at $x=10^{-3}$. However, as discussed earlier, Eq.~(\ref{undens}) is not always obeyed by the KMRW UPDFs, in particular at large $x$, see also Ref. \cite{Valeshabadi:2021smo}. In a variable-flavor-number scheme, the two main contributions to $D$-meson production are given by the $gg\to c\bar{c}$ and $cg\to cg$ processes. While good results have been recently obtained with a modified version of the strong-ordering KMRW UPDFs \cite{Guiot:2021vnp}, we demonstrate in Fig.~\ref{cgcg} that with the AO KMRW UPDFs the $cg\to cg$ process already overestimates the $D$-meson cross section.
The calculation is performed with the \KaTie\ event generator \cite{katie} with the same setup as Ref.~\cite{Guiot:2021vnp}. We used the Peterson fragmentation function \cite{Peterson:1982ak} with $\epsilon_c=0.05$.\\

To summarize, the issues of the KMRW formalism are
\begin{enumerate}
    \item Sudakov factor larger than $1$.
    \item UPDFs unconstrained by Eq.~(\ref{undens}) in the region $k_t\in[\mu,\infty]$.
    \item Overestimation of $D$- and $B$-meson cross section (only for the AO cutoff).
    \item Divergences associated with the use of the light-cone gauge We will discuss this point in Sec. \ref{secmkmrw}.
    \item Step behavior at $k_t=\mu_0$ due to Eq.~(\ref{conkt}). 
    \item Nonequivalence of the differential and integral definitions (and its consequences, e.g., the integral definition does not have exact normalization).
\end{enumerate}

In Sec. \ref{secsol}, we review quickly some solutions to the last two issues listed above. In Sec. \ref{secmkmrw}, we argue and demonstrate with an explicit example that a modification of the normalization condition solves all of them.\\

A clue that the third point of the above list can be solved by  changing the normalization can be found in Fig.~\ref{wmrpdf}. Here, the dotted lines correspond to the AO KMRW UPDFs of Ref.~\cite{wmr} integrated up to infinity. The observation that they overshoot the CTEQ14 PDFs explains the overestimation of the $D$-meson cross section. Having the UPDFs integrated up to infinity to agree with the collinear PDFs should solve this issue.\\

\section{Proposed solutions to the non-equivalence of the differential and integral definitions\label{secsol}}

The main goal of the first paper \cite{Valeshabadi:2021smo} presented in this section is to provide a solution to point 6 of the above list. The second paper focuses on Drell-Yan data. It proposes a modification of the KMRW formalism, solving points 5 and 6. 

\subsection{Cutoff-dependent PDFs\label{seccpdf}}
We have seen that the differential and integral definitions are not equivalent if $z_{ab}^{\text{max}}$ is not sufficiently close to 1. The authors of Ref.~\cite{BGS} proposed to use cutoff-dependent PDFs in Eq.~(\ref{def1}). In that case, the use of the cutoff-dependent DGLAP equation is justified for all $z_{ab}^{\text{max}}$. More recently, the authors of Ref.~\cite{Valeshabadi:2021smo} showed that a good numerical agreement between the differential and integral definitions is obtained if one combines the conditions of Refs. \cite{BGS,Guiot:2019vsm}, i.e., cutoff-dependent PDFs, and Eq.~(\ref{def3}) instead of Eq.~(\ref{def2}).\footnote{While we found Ref. \cite{Valeshabadi:2021smo} interesting, we do not fully agree with the claim that the only difference between the KMR \cite{kmr,kimthe} and Watt-Martin-Ryskin (WMR) \cite{wmr} approaches is the use of the cutoffs. The differences between these two formalism are discussed in the annex.} In that case, the UPDFs obey the normalization condition
\begin{equation}
    \widetilde{f}_a(x,\mu, z_{ab}^{\text{max}})=\int^{\mu^2}_0 F_a(x,k_t;\mu)dk_t^2, \label{normcut}
\end{equation}
with $\widetilde{f}_a(x,\mu, z_{ab}^{\text{max}})$ the cutoff-dependent PDFs. Here, $z_{ab}^{\text{max}}$ can take any value between 0 and 1. The usefulness of these functions is not clear, and we have
\begin{equation}
    \int_0^1 dx\widetilde{f}_{d_\text{val}}(x,\mu, z_{ab}^{\text{max}})\neq 1
\end{equation}
for the proton. In our opinion, the use of cutoff-dependent PDFs is not satisfying.

\subsection{$x$-dependent Sudakov factor\label{secmax}}
Nefedov and Saleev proposed a modified version of the KMRW UPDFs. In \cite{Nefedov:2020ugj}, the UPDFs obey exactly Eq.~(\ref{undens}) but Eq.~(\ref{conkt}) for $k_t<\mu_0$ is apparently not used\footnote{Here, we comment section III of Ref.~\cite{Nefedov:2020ugj}. Later, the definition of the UPDFs is modified further, but this modification is irrelevant to present discussion.}. In this case, we see from the differential definition, Eq.~(\ref{def1}), that the Sudakov factor should obey $T_a(k_t=\mu,\mu,x)=1$ and $T_a(k_t=0,\mu,x)=0$. We have anticipated that $T_a$ will depend on the longitudinal momentum fraction $x$. The method used by the authors of Ref.~\cite{Nefedov:2020ugj} is to fix the integral definition
\begin{equation}
F_a(x,k_t;\mu)=\frac{\alpha_s(k_t^2)}{2\pi}\frac{T_a(k_t,\mu,x)}{k_t^2}\sum_{b}\int_x^{\Delta(k_t,\mu)}dz\hat{P}_{ab}(z,k_t)\widetilde{f}_b\left(\frac{x}{z},k_t\right),\label{def5}
\end{equation}
and ask the exact equivalence with the differential form Eq.~(\ref{def1}) [with $T_a(k_t,\mu)$ replaced by $T_a(k_t,\mu,x)$]. They obtained a differential equation for the Sudakov factor with the solution \cite{Nefedov:2020ugj}
\begin{equation}
    T_a(k_t,\mu,x)=\exp\left(-\int_{k_t^2}^{\mu^2}\frac{dq^2}{q^2}\frac{\alpha_s(q^2)}{2\pi}(\tau_a(q^2,\mu^2)+\Delta \tau_a(q^2,\mu^2,x)) \right),\label{sudax}
\end{equation}
where
\begin{align}
    \tau_a(q^2,\mu^2)&=\sum_b\int_0^{\Delta(q,\mu)}dz z P_{ba}(z),\\
    \Delta\tau_a(q^2,\mu^2,x)&=\sum_b\int_{\Delta(q,\mu)}^1 dz\left(zP_{ba}(z)-\frac{\widetilde{f}_a\left(\frac{x}{z},q\right)}{\widetilde{f}_a(x,q)}P_{ab}(z)\Theta(z-x) \right).
\end{align}
Several comments are in order. They used the exact DGLAP equation so their result is valid for all $\Delta(k_t,\mu)$. For practical use, an equation similar to Eq.~(\ref{conkt}) is still required. Indeed, both Eqs.~(\ref{def5}) and (\ref{sudax}) depend on $\widetilde{f}_a(x,k_t)$, unknown for $k_t<\mu_0$. Finally, nothing prevents the Sudakov factor defined in Eq.~(\ref{sudax}) from being larger than 1 for $k_t>\mu$. We cannot use the trick of Eq.~(\ref{sudamod}) because a modification of Eq.~(\ref{sudax}) would break the equivalence of the differential and integral definitions. Note, however, that we could stop interpreting the Sudakov factor Eq.~(\ref{sudax}) as a probability, and interpret it as a function with the appropriate behavior to make the differential and integral form of the UPDFs equivalent.

\section{Interlude: The parton-branching approach\label{secpb}}
It is interesting to look at the parton-branching (PB) approach \cite{Hautmann:2017fcj,Hautmann:2017xtx}, where UPDFs obey the normalization condition:
\begin{equation}
    \widetilde{f}_a(x,\mu)=\int^{\infty}_0 \frac{F_a(x,\bm{k};\mu)}{\pi}d^2\bm{k}. \label{norinf}
\end{equation}
Integrating up to infinity solves several issues. For instance, the normalization condition (\ref{undens}) does not constrain the UPDFs in the region $k_t>\mu$ \cite{Guiot:2019vsm}, see also \cite{Hautmann:2019biw,Guiot:2018kfy}. However, in the calculation of the cross section, the $k_t$ integration is done up to values much larger than $\mu$. In other words, the unconstrained part of the UPDFs contributes to the cross section. Then, the overestimation of the $D$-meson cross section by the AO KMRW UPDFs is not necessarily a surprise, see Fig. \ref{cgcg}.\\

In the PB approach, the cutoff-dependent DGLAP equation is given by Eq.~(\ref{wmrdglap}), with $z_{ab}^{\text{min}}=0$ and $z_{ab}^{\text{max}}=z_M$. UPDFs obeying Eq.~(\ref{norinf}) are obtained by solving an integro-differential equation
\begin{multline}
\frac{\partial F_a(x,\bm{k};\mu;z_M)}{\partial \ln\mu^2}=\sum_{b}\frac{\alpha_s(\mu^2)}{2\pi}\left[\int_x^{z_M(\mu)}\hat{P}_{ab}(z)F_b\left(\frac{x}{z},\bm{k}+\bm{k'}(\mu,z);\mu;z_M\right)dz \right.\\
\left. -F_a(x,\bm{k};\mu;z_M)\int^{z_M(\mu)}_0 z\hat{P}_{ba}(z)dz \right],\label{eqdpb}
\end{multline}
which, once integrated over $k_t$, gives the cutoff-dependent DGLAP equation. The cutoff-dependent and exact DGLAP equations are equivalent, up to correction of order $\mathcal{O}(1-z_M)$ \cite{Hautmann:2017fcj}, and one should consequently choose $z_M$ close to 1.

\subsection{Comment on the $z_M$ dependence of the parton-branching UPDFs}
The authors of Ref.~\cite{Hautmann:2017fcj} observed that with the angular-ordering relation,
\begin{equation}
    \bm{k'}=(1-z)\bm{\mu}, \label{aorel}
\end{equation}
the UPDFs are independent of $z_M$, for $z_M$ in the region $(1-z_M)\sim \mathcal{O}(\Lambda_{QCD}/\mu)$. It is not the case with the transverse-momentum ordering
\begin{equation}
    \bm{k'}=\bm{\mu}.
\end{equation}
 This observation can be simply explained (here and in the following, we take $z_M$ constant). We start by studying the $z_M$ dependence of the collinear PDFs $\widetilde{f}_a(x,\mu,z_M)$ by taking the derivative of Eq.~(\ref{wmrdglap})
\begin{equation}
\frac{\partial \widetilde{f}_a(x,\mu,z_M)}{\partial \ln\mu^2\partial z_M }=\sum_{b}\frac{\alpha_s(\mu^2)}{2\pi}\left[\hat{P}_{ab}(z_M)\widetilde{f}_b\left(\frac{x}{z_M},\mu\right)dz-\widetilde{f}_a(x,\mu)z_M\hat{P}_{ba}(z_M)dz \right].
\end{equation}
Consider the case $a=g$. At $z_M=1$, the two terms involving $\hat{P}_{gg}$ cancel exactly. Using the large $z$ expansion
\begin{equation}
    \widetilde{f}_b\left(\frac{x}{z},\mu\right)=\widetilde{f}_b(x,\mu)+(1-z)\frac{\partial \widetilde{f}_b(x,\mu)}{\partial \ln x}+\mathcal{O}(1-z)^2,
\end{equation}
we find
\begin{multline}
\frac{2\pi}{\alpha_s(\mu^2)}\frac{\partial \widetilde{f}_g(x,\mu,z_M)}{\partial \ln\mu^2\partial z_M}=(1-z_M)\hat{P}_{gg}(z_M)\left(\widetilde{f}_g(x,\mu)+\frac{\partial \widetilde{f}_g(x,\mu)}{\partial \ln x}\right)\\
+\sum_{i=q,\bar{q}}\hat{P}_{gi}(z_M)\widetilde{f}_i \left(\frac{x}{z_M},\mu\right)-z_M\hat{P}_{ig}(z_M)\widetilde{f}_g(x,\mu)\label{depzm}
\end{multline}
The factor $(1-z_M)$ cancels the divergence in $\hat{P}_{gg}$. Thus, the rhs of Eq.~(\ref{depzm}) is finite. Consequently, a small variation of $z_M$ in the region $(1-z_M)~\sim \mathcal{O}(\Lambda_{QCD}/\mu)$ implies a small change for $\widetilde{f}_a$. A similar conclusion is obtained for $\widetilde{f}_q$.\\

Following the same procedure for Eq.~(\ref{eqdpb}), we find
\begin{multline}
\frac{2\pi}{\alpha_s(\mu^2)}\frac{\partial F_g(x,\bm{k};\mu;z_M)}{\partial \ln\mu^2\partial z_M}=\hat{P}_{gg}(z_M)\left(F_g(x,\bm{k}+\bm{k'};\mu;z_M)\frac{}{}\right.\\
\left. +(1-z_M)\frac{\partial F_g(x,\bm{k}+\bm{k'};\mu;z_M)}{\partial \ln x}\right)-z_M\hat{P}_{gg}(z_M)F_g(x,\bm{k};\mu;z_M)
+...\label{expF}
\end{multline}
where the dots include the finite contributions from $\hat{P}_{ig}$ and $\hat{P}_{gi}$. We observe that, because of $\bm{k'}$, it is not possible anymore to factorize $(1-z_M)$ in front of $F_g$, and the divergence in $\hat{P}_{gg}$ is not canceled. It is the infrared divergence discussed in \cite{Hautmann:2007uw}. While it is regularized with a subtraction method in \cite{Hautmann:2007uw}, the strategy of the PB approach is to make a specific choice for $\bm{k'}$. Using
\begin{equation}
    \bm{k'}=(1-z)^d\bm{A},
\end{equation}
with $d\geq1$ and $\bm{A}$ a finite two-dimensional vector, and doing an expansion of $F_g(x,\bm{k}+\bm{k'};\mu;z_M)$ around $\bm{k}$, we see that the rhs of Eq.~(\ref{expF}) is free of divergences
\begin{multline}
\frac{2\pi}{\alpha_s(\mu^2)}\frac{\partial F_g(x,\bm{k};\mu;z_M)}{\partial \ln\mu^2\partial z_M}=(1-z_M)\hat{P}_{gg}(z_M)\left(F_g(x,\bm{k};\mu;z_M)\frac{}{}\right.\\
\left. +(1-z_M)^{d-1} \bm{A}.\bm{\nabla}_{k}F_g(x,\bm{k};\mu;z_M)+\frac{\partial F_g(x,\bm{k}+\bm{k'};\mu;z_M)}{\partial \ln x}\right)
+... \label{derpb}
\end{multline}
While the angular-ordering condition has the appropriate form, with $d=1$ and $\bm{A}=\bm{\mu}$, the transverse-momentum condition does not, explaining the observed dependence of the UPDFs with $z_M$.

\section{Modified KMRW UPDFs \label{secmkmrw}}

We gave some phenomenological arguments in favor of the normalization condition (\ref{norinf}), in particular, those related to the $D$-meson cross section. Another clue is that the PB formalism does not encounter the issues listed in Sec.~\ref{seciss}. Finally, we remind that while the PDFs UV divergences in the integral over all $k_T$ can be managed with a cutoff, as in Eq.~(\ref{undens}), the preferred choice is the use of renormalization. In that case, the integration over all $k_T$ is kept, and the UV divergences are subtracted. The second choice is generally preferred because the PDFs defined with a cutoff have divergences related to the use of the light-cone gauge, see \cite{Collins:2003fm,col,Aslan:2022zkz} for an extended discussion. Consequently, we build a modified version of the AO KMRW UPDFs with normalization (\ref{norinf}) and show that none of the issues mentioned in Sec.~\ref{seciss} are present. We will refer to these distributions as mKMRW UPDFs.

\subsection{Differential and integral definitions \label{seckmrlike}}
 We start with the differential definition 
\begin{equation}
    F_a(x,k_t;\mu)=\frac{\partial}{\partial k_t^2}\left[T_a(k_t;\mu)\widetilde{f}_a(x,\mu)\right],\label{wmrlike}
\end{equation}
where the argument of the collinear PDFs is $\mu$ rather than $k_t$. This is necessary to avoid the unknown and unphysical values $\widetilde{f}_a(x,\infty)$ and $\widetilde{f}_a(x,0)$. This modification is also required by the DGLAP equation, see Sec.~\ref{secdglap}. The reason for changing the comma to a semicolon in $T_a(k_t;\mu)$ is explained after Eq.~(\ref{aocut}). With the normalization (\ref{norinf}), the Sudakov factor should obey
\begin{equation}
    T_a(\infty;\mu)-T_a(0;\mu)=1. \label{cd1}
\end{equation}
The minimal modification of the Sudakov factor is
\begin{equation}
T_a(k_t;\mu)=\exp\left(- \int_{k_t^2}^{\infty}\frac{dq^2}{q^2}\frac{\alpha_s(q^2)}{2\pi}\sum_b\int_0^{\Delta(q,\mu)}dz\, z \hat{P}_{ba}(z)\right),\label{sudinf}
\end{equation}
with the AO cutoff
\begin{equation}
    \Delta(q,\mu)=\frac{\mu}{\mu+q}.\label{aocut}
\end{equation}
The main change is the replacement of $\mu^2$ by $\infty$ for the upper limit of integration, similar to the change made in the normalization condition, going from Eq.~(\ref{undens}) to Eq.~(\ref{norinf}). A secondary but necessary change is that a cutoff is used for all splitting functions. In principle, the cutoff could depend on the splitting function, but for simplicity, we always use the same expression Eq.~(\ref{aocut}). Note that Eq.~(\ref{sudinf}) solves the issue of Sudakov factors larger than 1. We change the notation in $T_a(k_t;\mu)$ to clarify that the integral runs between $k_t^2$ and $\infty$. However, this function still depends on $\mu$ through the AO cutoff. 

The integral on $z$ can be done analytically, leading to
\begin{align}
    \int_0^\Delta dz\, z \hat{P}_{qq}&=-C_F\left(\frac{\Delta^3}{3}+\frac{\Delta^2}{2}+2[\Delta+\ln(1-\Delta)]\right), \label{pqqint}\\
    \int_0^\Delta dz\, z \hat{P}_{gq}&=C_F\left(2\Delta-\Delta^2+\frac{\Delta^3}{3}\right),\\
    \int_0^\Delta dz\, z \hat{P}_{qg}&= T_R\left(\frac{2\Delta^3}{3}-\Delta^2+\Delta \right),\\
    \int_0^\Delta dz\, z \hat{P}_{gg}&=-2C_A\left(\Delta^2-\frac{\Delta^3}{3}+\frac{\Delta^4}{4}+\ln(1-\Delta) \right).
\end{align}

\begin{figure}[h]
 \hspace*{-2cm}\includegraphics[width=38pc]{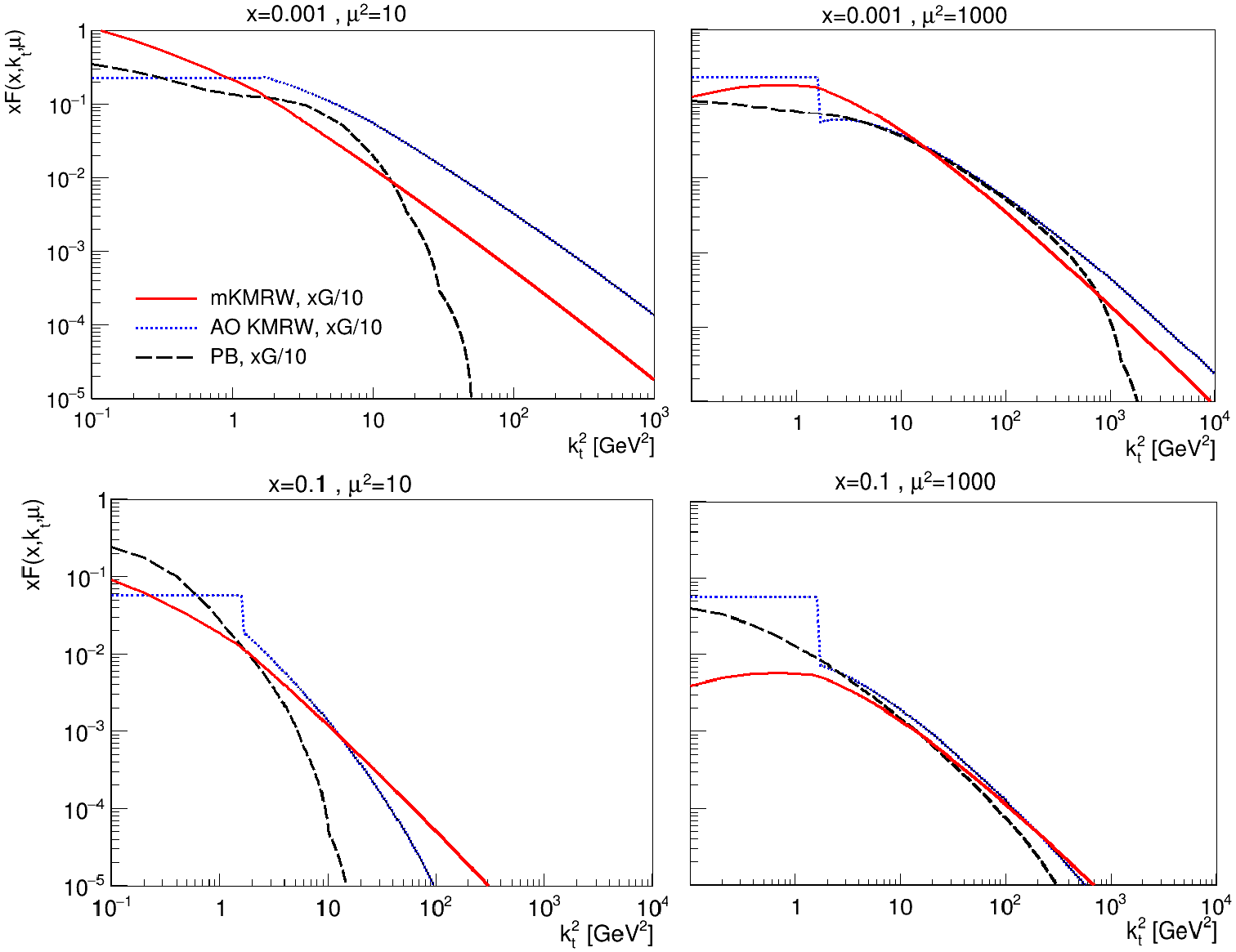}
\caption{\label{toypdf} Comparison of PB (dashed line) and AO KMRW unintegrated gluons (dotted line) with Eq.~(\ref{mkmrw}) (full line).}
\end{figure}
\begin{figure}[h]
\begin{center}
 \includegraphics[width=21pc]{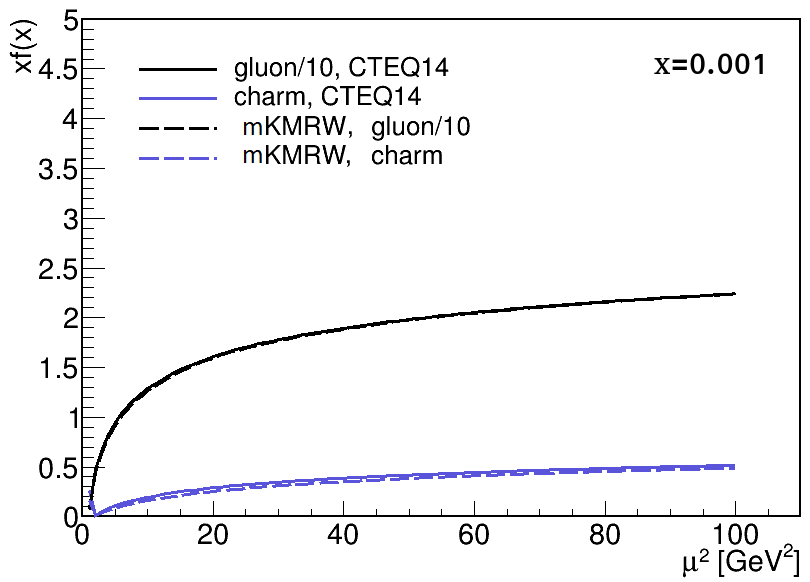}
\end{center}
\caption{\label{intnew} Comparison of CT14 PDFs (full lines) with integrated mKMRW UPDFs (dashed lines) based on Eq.~(\ref{norinf}).}
\end{figure}

We see that the Sudakov factor has the desired property, Eq.~(\ref{cd1}). Indeed, $T_a(\infty,\mu)=1$ and $T_a(0,\mu)=0$. Moreover, the Sudakov factor is larger than 0 for $0<k_t<\infty$. This is true thanks to the cutoff $\Delta$ which brings powers of $(\mu+q)^{-n}$ with $n>0$. For instance, in the limit $q\gg \mu$, and using the Taylor expansion of the logarithm, Eq.~(\ref{pqqint}) gives $C_F\mu^2/(2q^2)$. Taking the derivative in Eq.~(\ref{wmrlike}), we find the integral form of the mKMRW UPDFs
\begin{equation}
    F_a(x,k_t;\mu)=\frac{\alpha_s(k_t^2)}{2\pi k_t^2}T_a(k_t;\mu)\widetilde{f}_a(x,\mu)\sum_b\int_0^{\Delta(k_t,\mu)} dz\, z \hat{P}_{ba}(z).\label{mkmrw}
\end{equation}
This equation has a simple interpretation in terms of parton branching. We discuss this point in detail in Sec. \ref{secpbint} and compare with the usual KMRW formalism. To avoid the Landau pole, we use a saturated form of the coupling constant \cite{Dokshitzer:2000yu}
\begin{equation}
    \alpha_s(Q^2)=\text{min}\left[\frac{12\pi}{(33-6)\ln\left(\frac{Q^2}{\lambda_{QCD}^2} \right)},0.4 \right].
\end{equation}
The result of Eq.~(\ref{mkmrw}) is compared to the AO KMRW and PB UPDFs in Fig.~\ref{toypdf}. As expected, we observe that the mKMRW result is generally below the AO KMRW line. Before discussing Fig.~\ref{toypdf} in more detail, we demonstrate that the mKMRW UPDFs have the correct normalization by plotting Eq.~(\ref{norinf}), see Fig.~\ref{intnew}.
We observe a perfect agreement between the full and dotted lines. An unusual (but not necessarily surprising) feature of Fig.~\ref{toypdf} is the maximum at $k_t^2\sim 1$ GeV$^2$ for $\mu^2=10^3$ GeV$^2$. Finally, we observe that the shape of the AO KMRW and mKMRW UPDFs is similar for $k_t^2>\mu_0^2$. Consequently, the shape is not related to the normalization condition but to the evolution equation.\\

\begin{figure}[h!]
\begin{center}
 \includegraphics[width=23pc]{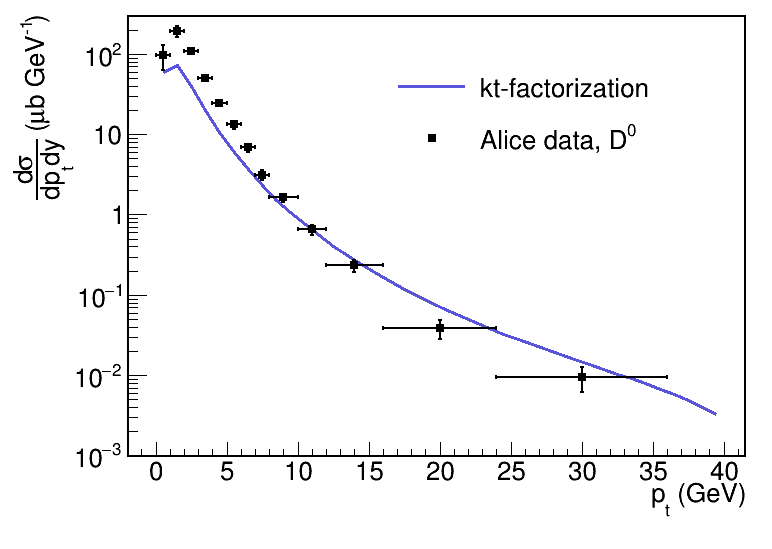}
\end{center}
\caption{\label{Dnew} $D$-meson production obtained with the mKMRW UPDFs and the event generator \KaTie\ . The calculation includes flavor-excitation and flavor-creation contributions ($gg\to c\bar{c}+q\bar{q}\to c\bar{c}+ac\to ac$, with $a$ any light parton).}
\end{figure}

In conclusion, the mKMRW UPDFs solve all the issues mentioned in Sec.~\ref{seciss}. In particular, the differential and integral definitions are equivalent, the UPDFs have exact normalization, and the $D$-meson cross section is no more overestimated by the $cg\to cg$ process. Indeed, in Fig.~\ref{Dnew} we show the result obtained including all flavor-excitation and flavor-creation contributions to $D$-meson production.
Note that the slight overestimation at $p_t\sim 30$ can easily be accommodated by a small increase of the fragmentation parameter $\epsilon_c$. We did not change this parameter to make the comparison with Fig.~\ref{cgcg} and Ref. \cite{Guiot:2021vnp} straightforward.

\subsection{DGLAP equation \label{secdglap}}

Taking the derivative of Eq.~(\ref{norinf}) with respect to $\mu^2$ and using the differential definition Eq.~(\ref{wmrlike}) gives
\begin{equation}
    \frac{d\widetilde{f}_a(x,\mu)}{d\mu^2}=\int_0^\infty \frac{\partial}{\partial k_t^2}\left[\frac{d}{d\mu^2}T_a(k_t;\mu)\widetilde{f}_a(x,\mu)+T_a(k_t;\mu)\frac{d}{d\mu^2}\widetilde{f}_a(x,\mu)\right]dk_t^2.
\end{equation}
Using Eq.~(\ref{cd1}), the second term gives the exact DGLAP equation. It is straightforward to show that the first term gives zero. We conclude that changing $k_t$ by $\mu$ in the differential definition, Eq.~(\ref{wmrlike}), is imposed by the DGLAP equation.\\

Note that we can use any scale in the cutoff of the Sudakov factor, for instance,
\begin{equation}
    F_a(x,k_t,Q;\mu)=\frac{\alpha_s(k_t^2)}{2\pi k_t^2}T_a(k_t;Q)\widetilde{f}_a(x,\mu)\sum_b\int_0^{\Delta(k_t,Q)} dz\, z \hat{P}_{ba}(z), \label{lla}
\end{equation}
with $Q$ the hard scale. It will not introduce a dependence on a new scale in the collinear PDFs since
\begin{equation}
    \frac{d}{dQ} \widetilde{f}_a(x,\mu)=\int_0^\infty \frac{\partial}{\partial k_t^2}\left[\frac{d}{dQ}T_a(k_t;Q)\widetilde{f}_a(x,\mu)\right]dk_t^2=0.
\end{equation}
The form given in Eq.~(\ref{lla}) is compatible with the leading-log approximation of the TMDs, with an example given by Eq. (13.85) of Ref. \cite{col} for fragmentation functions. In particular, the Sudakov factor is a function of $k_t$ and $Q$, while the dependence on the factorization scale is entirely in the collinear functions.

\subsection{Comparison with the Collins-Soper-Sterman formalism}
In \cite{Nefedov:2020ugj}, the authors showed that the KMRW Sudakov factor gives the typical (double) logarithms of the Collins-Soper-Sterman (CSS) formalism \cite{css} with the correct leading-log coefficients. This conclusion is unchanged for the modified Sudakov factor (\ref{sudinf}) since the large logarithms rise from the lower bound of integration. We perform the explicit calculation for the quark Sudakov factor where
\begin{equation}
    \sum_b\int_0^{\Delta(q,\mu)}dz\, z \hat{P}_{bq}=-C_F\left(\frac{3}{2}\Delta^2 + 2\ln(1-\Delta) \right).
\end{equation}
The integral on $q$ can be performed exactly with the term proportional to $\Delta^2$ given by
\begin{align}
 \frac{\alpha_s}{2\pi}\frac{3}{2}C_F\int_{k_t^2}^{\infty}\frac{dq^2}{q^2}\left(\frac{\mu}{\mu+q}\right)^2&=\frac{\alpha_s}{\pi}\frac{3}{2}C_F\left[-x-\ln(1-x)\right]^{\frac{\mu}{\mu+k_t}}_0 \label{css1}\\
 &=\frac{3\alpha_s}{4\pi}C_F\left(\ln\left(\frac{(\mu+k_t)^2}{k_t^2}\right)-\frac{2\mu}{\mu+k_t}\right).
\end{align}
For $k_t\ll \mu$, this contribution is dominated by $\ln(\mu^2/k_t^2)$ with the coefficient found in \cite{Nefedov:2020ugj}. Integrating up to $\mu^2$ instead of $\infty$ leads to the replacement $0 \to  1/2$ in Eq.~(\ref{css1}). We see that the exponent of the usual and modified Sudakov factor differs only by a finite term, while the logarithmic structure is untouched.\\

The integration of the term proportional to $\ln(1-\Delta)$ reads
\begin{align}
\frac{\alpha_s}{\pi}C_F\int_{k_t^2}^{\infty}\frac{dq^2}{q^2}\ln\left(\frac{q}{q+\mu} \right)&=\frac{2\alpha_S}{\pi}C_F \left[\text{Li}_2(1-x)+\frac{\ln^2(x)}{2}\right]^1_{\frac{k_t}{k_t+\mu}}\\
&=-\frac{\alpha_s}{4\pi}C_F\ln^2\left(\frac{(k_t+\mu)^2}{k_t^2} \right) + \text{finite terms}.
\end{align}
The coefficients of the single- and double-logarithmic terms coincide with those of the CSS formalism.\footnote{The KMR Sudakov factor has to be compared with the square root of the CSS Sudakov factor.}

\subsection{Parton-branching interpretation \label{secpbint}}
In the KMRW formalism, UPDFs are built in two steps. The idea was to use a single-scale evolution equation, and introduce the second scale $\mu$ at the last step of evolution through the Sudakov factor $T_a(k_t,\mu)$. The interest is a formalism simpler than those for realistic two-scale evolution. In Ref. \cite{kmr}, the two steps are
\begin{enumerate}
    \item Evolution up to scale $k_t$. At this point, the parton transverse momentum is also $k_t$.
    \item No splitting between scales $k_t$ and $\mu$ implemented with the Sudakov factor $T_a(k_t,\mu)$.
\end{enumerate}
The main issue is that this reasoning assumes $k_t<\mu$. However, $k_t$ can be infinite or at least reach large values such as $\sqrt{s}$, see Eq.~(\ref{ktfac}). Then, $T_a(k_t,\mu)$ does not prevent new emissions between $\mu$ and $\infty$. Instead, the modified Sudakov factor presented in Eq.~(\ref{sudinf}) does. This inconsistency is directly related to some of the issues mentioned in this manuscript, e.g., to the Sudakov factor larger than 1. It can be solved by changing the usual KMRW normalization condition by Eq.~(\ref{norinf}).\\

\noindent With the mKMRW UPDFs, steps 1 and 2 are reversed:
\begin{enumerate}
    \item An initial parton evolves collinearly with the DGLAP equation up to scale $\mu$.
    \item Then, a single splitting generates the transverse momentum $k_t$.
\end{enumerate}
The initial-parton number density at scale $\mu$ is given by $\widetilde{f}(x,\mu)$. The probability for a splitting with transverse momentum $k_t$ is $\frac{\alpha_s}{2\pi k_t^2}\int dz \, P(z)$. Finally, the condition of a single splitting is imposed by the probability of no emission between scales $k_t$ and $\infty$, i.e., by the Sudakov factor Eq.~(\ref{sudinf}). These three factors together lead to Eq.~(\ref{mkmrw}).\footnote{Note, however, that we kept the factor $z$ in $\int dz \, zP(z)$ introduced in Ref. \cite{wmr}. It is not necessary, but our goal was to propose a modification as close as possible of the Watt-Martin-Ryskin paper.}\\

Some additional comments on Eq.~(\ref{mkmrw}) are in order. In principle, the final splitting, step 2, should change the values of both $x$ and $\mu$. Note, however, that in the perturbative region, TMD PDFs are related to collinear PDFs by \cite{Collins:2015rpa}
\begin{equation}
    F_{j/h}(x,b_t;\zeta;\mu)=\sum_k C_{j/k}(x/z,b_t;\zeta,\mu)\otimes f(z,\mu)\frac{dz}{z},
\end{equation}
where $b_t$ is the conjugate variable of $k_t$ and $\zeta$ a scale irrelevant for the present discussion. The coefficients $C_{j/k}$ are perturbatively calculable, with the leading order proportional to $\delta_{jk}\delta(x/z-1)$. To leading order, having $x$ on the rhs and lhs of Eq. (\ref{mkmrw}) is then not surprising. It can be explained by the fact that for $k_t<\mu$, the factor $\int^{\Delta(k_t,\mu)} z P(z)dz$ is dominated by large $z$.

Something similar happens to the factorization scale. In principle, the argument of the UPDFs should be $\mu^2+\delta \mu^2$, but a bit of kinematics shows that $\delta \mu^2 \ll \mu^2$ for $k_t\ll \mu^2$.\\

Eq.~(\ref{mkmrw}) is then consistent, but accurate mainly in the region $k_t<\mu$. The last point is also true for the usual KMRW UPDFs. We have seen that they have been built assuming $k_t<\mu$, and that the normalization condition (\ref{undens}) does not constrain the distribution in the region $k_t>\mu$. In other words, none of these distributions should be taken too seriously at large $k_t$. It is interesting to note that realistic two-scale evolution equations, like those used by the PB and CCFM formalisms, lead to a fast decrease of the distribution with $k_t$ in the region $k_t>\mu$; see \cite{Golec-Biernat:2019scr} for a recent work with the CCFM-K formalism.\\

The main difference between the usual and modified KMRW UPDFs is not the behavior in the region $k_t>\mu$ (see Fig. \ref{toypdf}), but the fact that the latter is free of all the issues mentioned in Sec. \ref{seciss}. It is directly related to the choice of normalization condition and Sudakov factor. The latter should prevent new emissions in the region $[k_t,\infty]$ and be smaller than 1. These conditions are fulfilled by Eq.~(\ref{sudinf}).

\subsection{Dependence of UPDFs with $k_t$}

A comparison of Fig.~\ref{Dnew} with the results of Ref. \cite{Guiot:2021vnp} shows that the PB and strong-ordering KMRW UPDFs give slightly better results for $D$-meson production than the mKMRW UPDFs. It could indicate that the $k_t$ dependence of these  functions should be preferred, in agreement with the discussion in Sec. \ref{secpbint}. However, we should be careful with this conclusion. Here, the main issue is the underestimation of experimental data at small transverse momentum. It is known from event generators that realistic calculations should include a spacelike cascade\footnote{The spacelike cascade refers to the perturbative evolution before the partonic hard scattering.} which gives a significant contribution precisely in this kinematical region. In principle, another observable giving information on the $k_t$ shape of UPDFs is the azimuthal correlations of $Q\bar{Q}$ pairs. Most studies focus on the azimuthal correlations of the heavy-quark pairs produced by the gluon fusion process, assuming this is the main contribution. In fact, in a variable-flavor-number scheme, the $cg\to cg$ process gives the main contribution at intermediate and large $p_t$, and a $\bar{c}$ is emitted by the spacelike cascade. Then, a realistic study of heavy-quark azimuthal correlations requires again the use of an event generator taking into account all flavors, even in the perturbative evolution. Recently, the authors of Ref.~\cite{Abdulhamid:2021xtt} showed that the azimuthal correlation between two jets is sensitive to the $k_t$ distribution.\\

\begin{figure}[h]
\begin{center}
 \includegraphics[width=21pc]{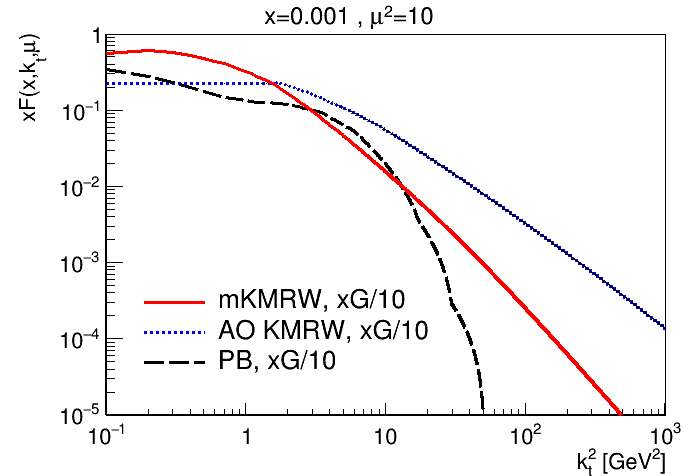}
\end{center}
\caption{\label{newdel} mKMRW UPDFs with cutoff Eq.~(\ref{del2}).}
\end{figure}
Concerning the mKMRW UPDFs, we can change the shape of the distribution by playing with the definition of the cutoff. For instance,
\begin{equation}
    \Delta(q,\mu)=\frac{\mu^2}{\mu^2+q^2} \label{del2}
\end{equation}
leads to the faster-decreasing distribution shown in Fig.~\ref{newdel}.
We did not explore which cutoff gives the best result for the $D$-meson cross section, as it is not the main focus of the present work.

\section{Conclusion}
The standard angular-ordering KMRW UPDFs present several issues: The differential and integral definitions are not equivalent, the normalization condition (\ref{undens}) is not exactly obeyed, and heavy-flavor production is overestimated. The last point has nothing to do with heavy quarks; the same will happen for the transverse-momentum distribution of any particle at intermediate $p_t$. The overestimation is due to the fact that the AO KMRW UPDFs are too large, because the tail of the distribution, at $k_t>\mu$, is not constrained by Eq.~(\ref{undens}).\\

In Sec.~\ref{secmkmrw}, we presented a modified version of the AO KMRW UPDFs, obeying normalization (\ref{norinf}). We showed that the differential and integral definitions are exactly equivalent, $T_a\leq 1$ for all $k_t$, and $D$-meson production is not overestimated anymore. The main effect of the change of normalization is to rescale the distribution, while the shape is not significantly affected. The definition proposed in Sec.~\ref{seckmrlike} is not unique, in particular because there is a freedom in the choice of the cutoff.\\

The main goal of the present work was not to build the best UPDFs, but to show that the normalization condition (\ref{undens}), together with the $k_t$-factorization hypothesis Eq.~(\ref{ktfac}), generates several issues. However, we showed that our proposition has the expected QCD properties of leading order high-energy calculations: the dependence on $\mu$ in the differential definition is imposed by the DGLAP equation, and the modified Sudakov factor has the same single and double logarithms as the CSS formalism. Finally, the x-dependence of the mKMRW and KMRW UPDFs is identical. It can be seen from Eqs. (\ref{def1}) and (\ref{wmrlike}), where the x-dependence of UPDFs is simply given by collinear PDFs.

\section{Annexe: Differences between the KMR and WMR formalisms}
The usual references for the KMR and WMR formalisms are \cite{kmr,wmr}. 
In \cite{Valeshabadi:2021smo}, the authors say that the only difference between these two formalisms is the implementation of the cutoff. This claim is not necessarily true since in \cite{kmr} the Sudakov factor is given by
\begin{equation}
    T_a(k_t,\mu)=\exp\left(- \int_{k_t^2}^{\mu^2}\frac{dq^2}{q^2}\frac{\alpha_s(q^2)}{2\pi}\sum_b\int_{0}^{1-\Delta} dz \, \hat{P}_{ba}(z) \right).\label{kmrta}
\end{equation}
A comparison with Eq.~(\ref{suda}) for WMR shows that the integrand is different due to the factor $z$ in front of the splitting function. We will call Eqs.~(\ref{suda}) and (\ref{kmrta}) the initial definitions. The final, or explicit, definitions for $T_q$ and $T_g$ are given in \cite{wmr}, but not in \cite{kmr}. However, explicit expressions for the KMR formalism can be found in \cite{kimthe}. Surprisingly, the integrands are now the same in both formalisms. But going from Eq.~(\ref{kmrta}) to final expressions in \cite{kimthe}, manipulations such as $\sum_a P_{aq}=P_{qq}$ (instead of $P_{qq}+P_{gq}$) or $\sum_a P_{ag}=zP_{gg}+n_fP_{qg}$ are found. Since, in the case of KMR, the initial and final expressions are not equivalent, saying that the integrand of the Sudakov factor is the same in both formalisms is then a matter of choice, depending on what we consider to be the true definitions (initial or final).

\section*{Acknowledgments}
We acknowledge support from ANID PIA/APOYO AFB180002 (Chile).


\end{document}